\documentclass[twocolumn,preprintnumbers,nofootinbib,amsmath,amssymb]{revtex4}
\usepackage{mathtext}
\usepackage{epsfig,amsfonts}% Include figure files
\usepackage{bm}% bold math

\newcommand\de{\delta}

\newcommand\De{\Delta}

\newcommand\Lam{\Lambda}

\newcommand\half{{\frac{1}{2}}}

\newcommand\lam{\lambda}

\newcommand\IM{{\rm Im}}

\begin{document}

\title{Single pole dominance in short- and intermediate-range
$\bm{NN}$ interaction}% Force line breaks with \\
\author{V.I. Kukulin}
\email{kukulin@nucl-th.sinp.msu.ru}
\author{V.N. Pomerantsev}
\email{pomeran@nucl-th.sinp.msu.ru}
\author{O.A. Rubtsova }%
\email{rubtsova-olga@yandex.ru}
\author{M.N. Platonova }%
\email{platonova@nucl-th.sinp.msu.ru}

\affiliation{%
Skobeltsyn Institute of Nuclear Physics, Lomonosov Moscow State
University, Leninskie Gory 1/2, 119991 Moscow, Russia}

\date{\today}% It is always \today, today,
\begin{abstract}
It is demonstrated for the first time that both elastic and
inelastic $NN$ scattering at laboratory energies up to 600--800
MeV, at least in some partial waves, can be described by a
superposition of the conventional long-range one-pion exchange and
a specific short-range interaction induced by the $s$-channel
dibaryon exchange. For the $^3P_2$, $^1D_2$ and $^3F_3$ partial
waves, the pole parameters giving the best fit of the real and
imaginary parts of the $NN$ phase shifts are consistent with the
parameters of the respective isovector dibaryon resonances found
experimentally. In the $^1S_0$ partial channel, the suggested
interaction gives two poles of the $S$-matrix
--- the well-known singlet deuteron and an excited dibaryon. On the
basis of the results presented, a conclusion is made about the
nature of the $NN$ interaction and its strong partial-wave
dependence.
\end{abstract}

%\pacs{03.65.Nk,21.45.-v,25.45.De}

\maketitle

\section{Introduction}
\label{intro} Nowadays the Effective Field Theory (EFT), or Chiral
Perturbation Theory (ChPT), is a dominating framework for the
quantitative description of the $NN$ interaction at low and
moderate energies \cite{Weinberg,Machl,Eppelbaum,Ordones}. In this
approach, a peripheral part of the $NN$ interaction is described
via a superposition of terms of the perturbation theory series
corresponding to the subsequent orders, i.e., the leading order
(LO), the next-to-leading order (NLO), the next-to-next-to-leading
order (N$^2$LO), etc., while the short-range contributions are
parameterized through the so-called contact terms which, according
to the general concept, should be independent of energy and the
expansion order. By construction, this general approach is valid
until the collision energies $T_{\rm lab} \simeq 350$~MeV, while
at higher energies it should be supplemented by an appropriate
theoretical model to describe the short-range components of the
$NN$ interaction and the short-range $NN$ correlations in nuclei.
For this purpose, one can consider the well-known quark model in
its various versions (see, e.g., \cite{Stancu}). However such a
hybrid approach inevitably leads to serious difficulties with
double counting because in quark models the gluon exchanges
between quarks are usually supplemented by meson ($\pi$ and
$\sigma$) exchanges, which immediately results in appearance of
not only short-range but also long-range meson-exchange forces
between nucleons.

At the moment in nuclear physics there is a wide class of
phenomena where one observes a close connection between
short-range correlations of nucleons in nuclei and some
distortions of quark momentum distributions in the deuteron,
$^3$He and other nuclei~\cite{Zhihong,Weinstein}. These phenomena
include the EMC effect, DIS observations, cumulative effects, etc.
It is evident that neither the traditional meson-exchange nor the
modern EFT approaches are relevant for a description of such
effects occurring at very high momentum transfers. On the other
hand, since all these phenomena are closely related to nuclear
force and nuclear structure at short distances, they should be
described within some general scheme which includes the correct
treatment of the $2N$ and $3N$ forces.

Therefore, for further progress in this area it would be highly
desirable to treat the short- and intermediate-range $NN$
interactions by using some QCD-motivated model which could
reproduce correctly the basic effects of the nucleon quark
structure in different $NN$ partial waves, however {\em without
addressing} all complexity of the multi-quark dynamics. At the
same time, such a model description of the short-range $NN$
dynamics should not modify the known interaction at long
distances.

In our opinion, suitable objects, which, on the one hand, can
reproduce the main features of six-quark dynamics and, on the
other hand, are strongly coupled to the hadronic ($NN$, $N\Delta$
and $\Delta\Delta$) channels at low and moderate energies are
dibaryon resonances which were predicted by Dyson and Xuong
in 1964, at the very beginning of quark era \cite{Dyson}. Just
recently, after many years of rejection, doubt, and contradictory
findings, a number of dibaryon resonances have been eventually
confirmed in the modern high-precision experiments
\cite{Adlarson11,Adlarson14,Komarov,Adlarson18} (see also the
recent review \cite{Clement}).

Dibaryon resonances are very attractive to describe the
short-range $NN$, $N\De$ and $\De\De$ forces not only due to their
six-quark structure but, first of all, because they are specific
relatively long-lived states in which six-quark dynamics should be
manifested most clearly. So, the goal of the present paper is to
demonstrate that the $NN$ interaction at least in particular
partial-wave channels can be described properly by a superposition
of the long-range meson-exchange potentials and one simple
$s$-channel exchange potential driven by a dibaryon intermediate
state.

For the particular $NN$ partial waves considered below (and for
many others) it is crucial to include the $NN \to N\Delta$ (and/or
$NN \to \Delta\Delta$) coupling. In turn, the $NN \to N\Delta$
transition may or may not be accompanied by a dibaryon formation.
While the latter type of coupling has been effectively included
into our model consideration through the resonance parameters (in
fact, the isovector dibaryon width is mainly due to the $D \to
N\Delta$ decay), the former (pure $t$-channel) coupling is
actually a background to the $s$-channel dibaryon generation which
should be included explicitly into the $NN$ interaction potential.
Though the background processes may affect significantly the
description of $NN$ inelasticities above the resonance energy,
they will not change our main results concerning the impact of the
basic dibaryon mechanism at lower energies. Thus we postpone the
consistent treatment of the $NN \to N\Delta$ coupling to the
future work.

It should be noted that the initial version of the dibaryon model
for $NN$ and $3N$ interactions developed in \cite{JPhys2001} (see
also \cite{AnnPhys2010} and references therein) provided quite
encouraging results in description of $NN$ scattering in various
partial waves. In particular, it allowed to reproduce real parts
of phase shifts in different $NN$ channels up to energies $T_{\rm
lab} = 500$--$600$ MeV. However, since then dibaryons have
achieved a more reliable experimental status, while the parameters
of the real dibaryon poles found in the work \cite{JPhys2001} were
never compared to those of the physical dibaryon resonances. So,
in the present paper we tried to incorporate into the initial
model the experimentally found dibaryon resonances together with
their empirical parameters to describe simultaneously the real and
imaginary parts of the $NN$ phase shifts in a broad energy range.
The results of this study would allow to judge about the true
applicability and merit of such a non-conventional description of
$NN$ interaction.

The structure of the paper is as follows. In Sec. \ref{form}, we
introduce the two-channel formalism (with one external and one
internal channel) which is used further to describe the $NN$
partial phase shifts. In Sec. \ref{res1} and \ref{res2}, we
present the results for the particular isovector $NN$ partial
waves and illustrate effectiveness of the dibaryon mechanism. We
summarize the results and conclude in Sec. \ref{concl}.

\section{$\bm{NN}$ scattering driven by a single state with a complex energy
in the internal channel}
\label{form}

We consider below a two-channel model with one complex pole in the
effective interaction potential. This model corresponds to the
physical pattern of the $NN$ scattering driven by the traditional
one-pion exchange in the external channel and by one state with a
complex eigenvalue (i.e., the ``bare'' dibaryon state) in the
internal channel. The complex energy of this state can be
interpreted as a consequence of different modes of its decay which
do not include the $NN$ mode.\footnote{The initial dibaryon state
with the real energy can be treated in terms of the field theory
as a ``bare'' dibaryon, while the coupled-channel dibaryon which
is able to decay into the $NN$ continuous spectrum can be
identified as a ``dressed'' dibaryon. In this sense, the dibaryon
with the complex energy can be called as a ``semi-dressed'' one.
However, to retain the unified notation, we will refer to the
initial dibaryon with the complex energy as a ``bare'' one.}

The total matrix Hamiltonian for such a two-channel problem has the form:
\begin{equation}
H=\left(
\begin{array}{cc}
h_{NN}& \lam_1|\phi\rangle\langle \alpha|\\
\lam_1 |\alpha\rangle \langle \phi|& E_D|\alpha\rangle \langle
\alpha|\\
\end{array}
\right), \label{Ham}
\end{equation}
where the external-channel Hamiltonian
$h_{NN}=h_{NN}^{(0)}+V_{NN}$ acts in the space of $NN$ variables
and includes the peripheral $NN$ interaction $V_{NN}$ which is
exhausted by the one-pion exchange potential (OPEP):
\begin{equation}
V_{NN}=
\frac{f_{\pi}^2}{m_{\pi}^2}\frac{1}{q^2+m_{\pi}^2}\left(\frac{\Lambda_{\pi
NN}^2-m_{\pi}^2}{\Lambda_{\pi
NN}^2+q^2}\right)^2({\bm\sigma}_1\cdot {\bf q})({\bm
\sigma}_2\cdot {\bf q})\frac{({\bm\tau}_1\cdot {\bm\tau}_2)}{3},
\label{Vope}
\end{equation}
where $m_{\pi} = (m_{\pi^0} + 2m_{\pi^{\pm}})/3$ is the averaged
pion mass and $\Lambda_{\pi NN}$ --- the high-momentum cutoff
parameter\footnote{In our calculations, the averaged pion-nucleon
constant $f_{\pi}^2/(4\pi)=0.075$ and the soft cutoff
$\Lambda_{\pi NN}=0.65$~GeV/$c$ are used.}.

In case of the single-pole model, the Hilbert space of the
internal channel is one-dimensional and therefore the internal
Hamiltonian is reduced to a single term with a complex eigenvalue
$E_D=E_0 -i\Gamma_{\rm inel}/2$, the imaginary part of which is
defined by the decay width of the dibaryon into all inelastic
channels.

To determine the form factor $|\phi \rangle$ of the transition
between the external ($NN$) and the internal (dibaryon) channels,
it is necessary to use some microscopic model that describes both
channels within a unified approach. In our previous works
\cite{JPhys2001,AnnPhys2010}, a dibaryon model for nuclear forces
was developed, in which a microscopic six-quark shell model in a
combination with the $^3P_0$ quark mechanism of pion production
was used to determine the transition amplitude between two
channels. Note that the transition form factor $|\phi\rangle$ is a
function of the relative coordinate $r$ (or the relative momentum)
in the $NN$ channel, and also depends on the spin, isospin,
orbital and total angular momenta of the two-nucleon state. Here
we use the same form of this function as in Ref.~\cite{JPhys2001},
i.e., the harmonic oscillator term:
\begin{equation}
\phi(r)=Nr^{l+1}\exp\left[-\half\left(\frac{r}{r_0}\right)^2\right],
\end{equation}
where $l$ is the orbital angular momentum in the two-nucleon
system, $N$ --- the normalization factor, and $r_0$ --- the scale
parameter.

Since the total Hamiltonian (\ref{Ham}) couples the internal
 channel to the $NN$ relative-motion channel, it is convenient to
 exclude the internal channel by the commonly-used method \cite{Feshb} and to deal
 with the $NN$ relative motion only. Then the resulted effective energy-dependent
 Hamiltonian in the $NN$ channel takes the form:
\begin{equation}
H_{\rm eff}(E)=h_{NN} +\frac{\lam_1^2|\phi\rangle \langle
\phi|}{E-E_D}. \label{Heff}
\end{equation}
Due to the fact that the basic term of the effective Hamiltonian
(\ref{Heff}) has a separable form, it is convenient to define
explicitly an additional $t$-matrix in the distorted-wave
representation:
\begin{equation}
t(E)=\frac{\lam_1^2|\phi\rangle\langle\phi|}{E+i0-E_D-J_1(E+i0)},
\label{te}
\end{equation}
where $J_1(Z)$ is the matrix element of the resolvent of the
external $NN$ Hamiltonian $g_{NN}(Z)\equiv [Z-h_{NN}]^{-1}$:
\begin{equation}
J_1(Z)=\lam_1^2\langle \phi|g_{NN}(Z)|\phi\rangle.
\end{equation}
Note that the imaginary part of this function at a real positive
energy can be found in an explicit form as
\begin{equation}
\IM \, J_1(E+i0)= -\pi \lam_1^2|\langle \phi|\psi_0(E)\rangle|^2,
\end{equation}
where $|\psi_0(E)\rangle$ is the scattering function for the
external Hamiltonian $h_{NN}$. Using the formula for the
transition operator (\ref{te}), one can easily obtain the
expression for the total $S$-matrix:
\begin{equation}
S(E)=e^{2i\de_0}\frac{E-E_D-J_1^*(E+i0) }{E-E_D- J_1(E+i0)},
\end{equation}
where $\delta_0(E)$ is the phase shift for the external Hamiltonian
$h_{NN}$. Thus, the pole of the total $S$-matrix can be found in a
complex energy $Z$ plane from the condition:
\begin{equation}
Z-E_D-J_1(Z)=0. \label{res_con}
\end{equation}
The condition (\ref{res_con}) makes it possible to find the
renormalized position of the dressed dibaryon resonance
$Z_R=E_R-i\Gamma_{\rm th}/2$ (relatively to the initial ``bare''
value of $E_D$), i.e., the complex function $J_1(Z)$ gives shifts
of the real and imaginary parts of the dibaryon pole. These shifts
arise due to coupling of the initial dibaryon to the external $NN$
channel. The invariant mass of the dressed dibaryon can be
calculated from
 the real part of the energy $E_R$ by the relation $M_{\rm
th}=2\sqrt{m(E_R + m)}$, where $m$ is the nucleon
mass\footnote{Here we use the ``minimal'' account of the
relativistic effects (see, e.g., Ref.~\cite{Geramb}) by keeping
the relation $E=k^2/m$ between the energy $E$ and the relative
momentum $k$ in the Lippmann--Schwinger equation, while
calculating $k$ from the laboratory energy $T_{\rm lab}$ via the
relation $k=\sqrt{m T_{\rm lab}/2}$. Finally, the invariant energy
is calculated from $E$ by using the formula
$\sqrt{s}=2\sqrt{m(E+m)}$.}.

For the effective account of inelastic processes and description
of the threshold behavior of the reaction cross section, one
should introduce the energy dependence of the bare dibaryon width
$\Gamma_{\rm inel}$. The main inelastic process for the isovector
$NN$ channels considered here is the one-pion production. The
three-body mode $D \to \pi N N$ dominates in the decay width of
dibaryons in these partial-wave channels, while the two-body mode
$D \to \pi d$ takes $\lesssim 30$\%~\cite{Strak91} of the total
dibaryon width and has the similar threshold behavior. Thus, the
respective inelastic width can be represented as follows:
\begin{equation}
\Gamma_D(\sqrt{s})=\left\{
\begin{array}{lr}
0,& \sqrt{s}\leq E_{\rm thr};\\\displaystyle
\Gamma_0\frac{F(\sqrt{s})}{F(M_0)},&\sqrt{s}>E_{\rm thr}\\
\end{array}\label{gamd}
\right.,
\end{equation}
where $\sqrt{s}$ is the total invariant energy of the decaying
resonance, $M_0$ --- the bare dibaryon mass, $E_{\rm
thr}=2m+m_\pi$ --- the threshold energy, and $\Gamma_0$ defines
the decay width at the resonance energy.

The function $F(\sqrt{s})$ should take into account the dibaryon
decay into the channel $\pi N N$. So that, for the given values of
the orbital angular momenta of the pion $l_{\pi}$ and $NN$ pair
$L_{NN}$, this function can be parameterized in the form:
\begin{equation}
F(\sqrt{s})=\frac{1}{s}\int_{2m}^{\sqrt{s}-m_{\pi}}dM_{NN}
\frac{q^{2l_\pi+1}k^{2L_{NN}+1}}{(q^2+\Lam^2)^{l_\pi+1}(k^2+\Lam^2)^{L_{NN}+1}},
\label{fpinn}
\end{equation}
where $\displaystyle
q={\sqrt{(s-m^2_\pi-M^2_{NN})^2-4m_\pi^2M_{NN}^2}}\Big/{2\sqrt{s}}$
is the pion momentum in the total center-of-mass frame,
$\displaystyle k=\half\sqrt{M_{NN}^2-4m^2}$ --- the momentum of
the nucleon in the center-of-mass frame of the final $NN$
subsystem with the invariant mass $M_{NN}$, and $\Lam$ --- the
high-momentum cutoff parameter which prevents an unphysical rise
of the width $\Gamma_{\rm inel}$ at high energies. The orbital
momenta $l_{\pi}$ and $L_{NN}$ may take different values however
their sum is restricted by the total angular momentum and parity
conservation. In practical calculations, the values of $l_{\pi}$,
$L_{NN}$ and $\Lam$ were adjusted to get the best description of
the phase shifts in the given $NN$ partial wave. It should be
emphasized that the values of these parameters affect mainly the
threshold behavior of the inelastic partial phase shifts. They are
used for the ``fine-tuning'' of the model while the resulting
phase shifts in a broad energy range are much more sensitive to
the mass and width of the bare resonance.

\begin{figure}[h!] \centering\epsfig{file=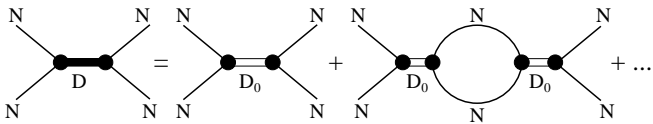,width=\columnwidth}
\caption{\label{diagram} The sum of terms corresponding to
dressing of the total dibaryon propagator. $D_0$ and $D$ are the
propagators of the ``bare'' and ``dressed'' dibaryons,
respectively.}
\end{figure}

Thus, we have formulated a simple model for coupling between the
external $NN$ (driven by OPEP) channel and the internal (``bare''
dibaryon) channel which leads to a renormalization of the complex
energy of the initial ``bare'' dibaryon and its transformation to
the real mass and width of the ``dressed'' dibaryon.  The simple
mechanism for coupling between the external and internal channels
can be represented graphically by the diagram series shown in
Fig.~\ref{diagram}. This series corresponds to the well-known
Dyson equation for a dressed particle in the quantum field theory.

\section{$\bm{NN}$ scattering in isovector channels}
\label{res1}

In this section, we consider as particular examples the isovector
partial waves ($^3P_2$, $^1D_2$ and $^3F_3$) of the $NN$
scattering, where the empirical phase shifts do not manifest
explicitly the behavior inherent to the repulsive core, at least
until relatively high collision energies $T_{\rm lab} \simeq
800$~MeV.\footnote{The fact that at higher energies $T_{\rm lab} >
800$~MeV the real phase shifts in some channels become negative
can be explained by strong absorption at these energies, i.e., by
appearance of a large imaginary part of phase shifts that can be
interpreted as a consequence of strong repulsion.} The resonance
behavior in these partial waves near the $N\De$ threshold was
established long ago in experiments \cite{Meshcher,Auer1,Auer2}
and was interpreted either by the threshold effect or the true
dibaryon formation. The dibaryon interpretation was then supported
by the partial-wave analyses of different groups which found the
$S$-matrix poles corresponding to dibaryon
resonances~\cite{Hoshiz,Arndt,Strak} (see also the review
paper~\cite{Strak91}). In our recent works \cite{NPA2016,PRD2016}
the importance of these three resonances was established for
description of the basic pion-production reaction $pp \to d
\pi^+$, and the $^3P_2$ dibaryon was shown to play a crucial role
in reproducing the polarization observables. Simultaneously, the
$^3P_2$ dibaryon was observed in the recent high-precision
experiment on the reaction $pp \to pp(^1S_0) \pi^0$
\cite{Komarov}.

It should be emphasized that previous analyses of $NN$ scattering
which considered dibaryon resonances generally suggested that the
dibaryon pole can give a significant contribution to the $NN$
phase shifts in a respective partial wave only near the resonance
energy. Contrary to this, we will show below that the single pole
(combined with a long-range OPEP contribution) can explain the
phase shifts behaviour in the isovector channels in a broad energy
range from zero up to the resonance energy. The isoscalar channels
of $NN$ scattering will be studied in our subsequent paper.

For description of the partial phase shifts, the Arndt
parametrization for the  $K$-matrix \cite{SAID2007,Geramb} is
used. For the uncoupled $NN$ channels, it has a simple form:
\begin{equation}
K=\tan\delta +i \tan^2\rho,
\end{equation}
where $\delta$ is the real phase shift and $\rho$ is a parameter
related to inelasticity. For the sake of simplicity, below we will
refer to the parameter $\rho$ as the imaginary phase shift.

\subsection{Channel $\bm{^3P_2}$}

The empirical $NN$ phase shifts in the triplet channel $^3P_2$ as
found by the George Washington University group (SAID)
\cite{SAID2007} do not display any sign of the repulsive core and
remain to be positive at least up to energies of $T_{\rm lab}
\simeq 1$ GeV.\footnote{Such a behavior can be explained in the
conventional $NN$ potential models by a complete compensation of
the short-range repulsive core by the very strong attractive
spin-orbital potential, so that, the resulting potential in this
channel turns out to be attractive \cite{Bohr}.} This can be
interpreted as a fact that the traditional repulsive core does not
play a crucial role in this channel, and thus $NN$ interaction
here is managed by a rather strong attraction, which is likely due
to generation of a dibaryon resonance.

Below we show that such an attraction can be reproduced by a
single dibaryon pole in the effective Hamiltonian (\ref{Heff}) via
varying its position $E_D=E_0-i\Gamma_D/2$ and the coupling
constant $\lam_1$ for the external and internal channels.

\begin{figure}[h!]
\centering\epsfig{file=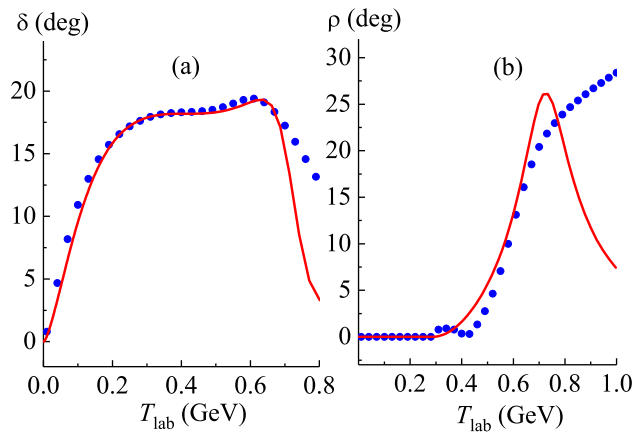,width=\columnwidth}
\caption{\label{fig2} Real $NN$ phase shifts (a) and parameters of
inelasticity (b) in the $^3P_2$ partial wave (solid curves) in
comparison with the SAID data \cite{SAID2007} (points).}
\end{figure}

The comparison of empirical (SAID) and theoretical $^3P_2$ $NN$
phase shifts in the energy interval $T_{\rm lab} =  0$--$800$~MeV
is presented in Fig.~\ref{fig2}.\footnote{Though the SAID results
for $np$ scattering are plotted in Fig.~\ref{fig2} and subsequent
figures, the $pp$ scattering phase shifts are indistinguishable
from them in the scale of the picture.} Here the following
potential parameters were used: $\lam_1=0.065$~GeV and
$r_0=0.71$~fm. The initial dibaryon mass was taken to be
$M_0=2.21$~GeV, and the inelastic width had the form (\ref{gamd}),
(\ref{fpinn}) with the parameters $\Gamma_0=0.096$~GeV, $l_\pi=2$,
$L_{NN}=0$ and $\Lam=0.3$ GeV/$c$.

It is seen from Fig.~\ref{fig2} that the single-pole model in a
combination with a simple OPEP provides almost quantitative
agreement with the empirical data for the real part and reasonable
agreement for the imaginary part of the $^3P_2$ phase shifts up to
energies $T_{\rm lab} \simeq 700$~MeV. Needless to say that the
above agreement for both real and imaginary parts of the phase
shifts, i.e., for elastic $NN$ scattering and meson production
simultaneously, was attained {\em with the same parameters} for
the bare dibaryon. Some discrepancy observed for the imaginary
part of the phase shift, starting just above the pion-production
threshold, is likely related to our simplified description of the
energy dependence of the dibaryon width, and can probably be
decreased by a more sophisticated treatment of $\Gamma_D(E)$. The
proper way to proceed here is to include explicitly in the model
the $N\Delta$ $P$-wave channels which couple to the $NN(^3P_2)$
channel.

Now, using Eq.~(\ref{res_con}), one can find easily the
renormalized position of the resonance pole in this channel. We
found the following parameters for the dressed dibaryon in the
channel $^3P_2$:  $M_{\rm th}(^3P_2)=2.23$ GeV and $ \Gamma_{\rm
th}(^3P_2)=0.15$ GeV. These parameters should be compared with the
respective experimental values found recently by the ANKE-COSY
Collaboration \cite{Komarov}: $M_{\rm exp}(^3P_2)=2.197(8)$ GeV
and $\Gamma_{\rm exp}(^3P_2)=0.130(21)$ GeV (the numbers in
parentheses denote the uncertainty in the last figure). It is
worth noting that the authors of \cite{Komarov} used two different
fitting procedures, the first leading to the above quoted values
and the second (the global fit) --- to somewhat larger values
$M_{\rm exp}(^3P_2)=2.207(12)$ GeV and $\Gamma_{\rm
exp}(^3P_2)=0.170(32)$ GeV still consistent with the above ones
within errors. Thus, the mass and width obtained in this work for
the $^3P_2$ dibaryon turn out to be rather close to their
experimental values (with account of the experimental
uncertainties).

Since our model does not include any other assumptions, except the
OPEP and presence of a single dibaryon pole, we owe to conclude
that the results obtained point directly to the dominance of the
intermediate dibaryon production in the $NN$ interaction in this
channel.

\subsection{Channel $\bm{^1D_2}$}
\begin{figure}[h!]
\centering\epsfig{file=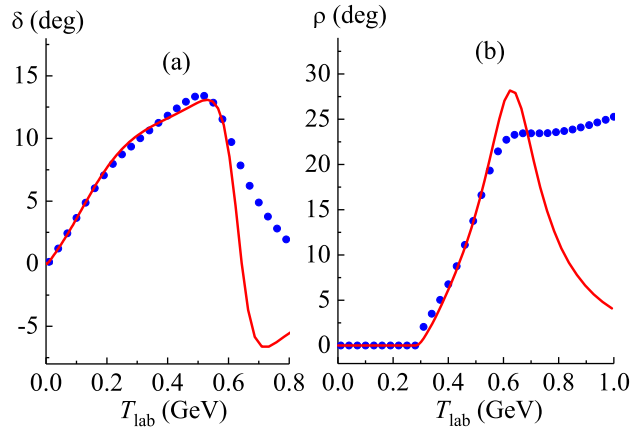,width=\columnwidth}
\caption{\label{fig3} The same as in Fig.~\ref{fig2} for the
$^1D_2$ partial wave.}
\end{figure}

A fully similar consideration of the $NN$ phase shifts in the
singlet channel $^1D_2$ leads to theoretical predictions shown in
comparison with the respective empirical data (SAID) in
Fig.~\ref{fig3}. The potential parameters for this channel have
been taken to be $\lam_1=0.048$~GeV and $r_0=0.82$~fm. The initial
dibaryon mass is $M_0=2.168$~GeV, and the width parameters are
$\Gamma_0=0.08$ GeV, $l_\pi=0$, $L_{NN}=1$ and $\Lam=0.2$ GeV/$c$.

Here we see again almost perfect agreement for the real and
reasonable agreement for the imaginary phase at energies up to
600~MeV. The parameters of the dressed dibaryon in the $^1D_2$
channel found in our calculations are $M_{\rm th}(^1D_2)=2.18$ GeV
and $\Gamma_{\rm th}(^1D_2)=0.11$ GeV. These parameters are
consistent with those found previously in
experiments~\cite{Meshcher,Auer2} and partial-wave
analyses~\cite{Hoshiz,Arndt,Strak} and turn out to be rather close
to the values obtained in the early experiment of 1955
\cite{Meshcher} where the $^1D_2$ resonance was first observed:
$M_{\rm exp}(^1D_2) \simeq 2.16$ GeV and $\Gamma_{\rm exp}(^1D_2)
\simeq 0.12$ GeV.

It is important to emphasize that the $NN$ channel $^1D_2$
strongly couples to the $N\Delta$ channel $^5S_2$, and just this
strong coupling with a subsequent decay of the $\Delta$-isobar
determines a large portion of inelasticity in the $^1D_2$ partial
wave. Since we did not take into account the $t$-channel $\Delta$
excitation in our model, the description of inelastic phase shifts
turns out to be not perfect. However, almost quantitative
description of the real part of the phase shifts until $T_{\rm
lab}=600$~MeV and reasonable description of their imaginary part
(with the same model parameters) point to the dominant role of the
dibaryon resonance (including its $N\Delta$ component) in this
channel as well.

\subsection{Channel $\bm{^3F_3}$}
First experimental evidence of the $^3F_3$ dibaryon resonance with
the mass $M_D \simeq 2.26$ GeV was obtained in 1977 \cite{Auer1}.
It was also established that there is a very large inelasticity in
the $^3F_3$ $NN$ partial wave. So, an interesting and challenging
question arises: to what degree the $^3F_3$ resonance contributes
to the real and imaginary parts of the $NN$ phase shifts in this
channel? From the first glance, the possible impact of the
dibaryon can be seen mainly at the energies close to the resonance
position. However, in general, the $s$-channel dibaryon exchange
in $NN$ scattering must be retraced also far from the resonance
energy and give an evident impact to the off-shell $t$-matrix.

\begin{figure}[h!]
\centering\epsfig{file=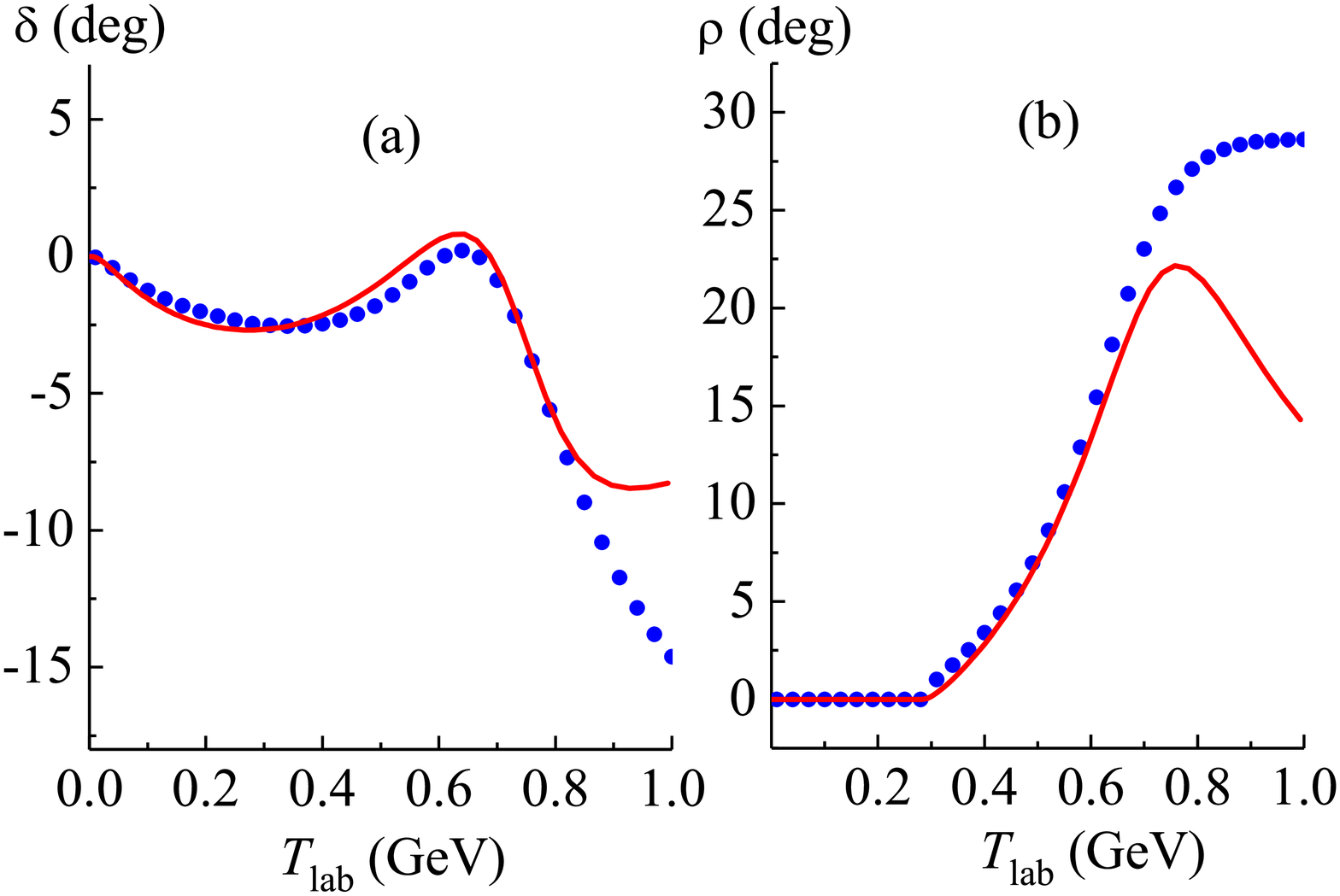,width=\columnwidth}
\caption{\label{fig4} The same as in Fig.~\ref{fig2} for the
$^3F_3$ partial wave.}
\end{figure}

In Fig.~\ref{fig4} the partial phase shifts in the $^3F_3$ channel
are compared with the empirical (SAID) data. As is clearly seen
from the figure, the single $^3F_3$ dibaryon pole (in a
combination with the peripheral OPEP) can reproduce very well the
real $NN$ phase shifts from zero energy up to 700 MeV (see
Fig.~\ref{fig4}). The parameters of the coupling potential in this
case are $\lam_1=0.06$ GeV and $r_0=0.5$ fm, while the pole
parameters are $M_0=2.23$ GeV and $\Gamma_0=0.15$ GeV with
$l_\pi=0$, $L_{NN}=2$ and $\Lam=0.1$ GeV/$c$.

The parameters of the dressed dibaryon found for the $^3F_3$
channel, i.e., $M_{\rm th}(^3F_3)=2.22$ GeV and $\Gamma_{\rm
th}(^3F_3)=0.17$ GeV, are also consistent with the previous
findings~\cite{Auer1,Auer2,Hoshiz,Arndt,Strak} and occur to be
rather close to the values obtained in the first experimental
observation of the $^3F_3$ resonance~\cite{Auer1}: $M_{\rm
exp}(^3F_3) \simeq 2.26$ GeV and $\Gamma_{\rm exp}(^3F_3) \simeq
0.2$ GeV.

The results presented in this section for the isovector dibaryon
resonances together with experimental data are summarized in
Table~\ref{Tab1}.
\begin{table}[h!]
\caption{Parameters of bare and dressed dibaryon resonances (in
GeV) for three isovector $NN$ channels in comparison with
experimental values taken from Refs.~\cite{Komarov} ($^3P_2$) and
\cite{Strak91} ($^1D_2$, $^3F_3$).\label{Tab1}}
\begin{center}\begin{tabular}{c c c c c c c}\hline ${}^{2S+1}L_J$& $M_0$&
$\Gamma_0$&$M_{\rm th}$&$\Gamma_{\rm th}$&$M_{\rm exp}$&$\Gamma_{\rm exp}$\\
\hline
$^3P_2$& 2.21 &0.096 &2.23&0.15& 2.197(8)&0.130(21) \\
$^1D_2$& 2.168 &0.08 &2.18&0.11& 2.14--2.18 &0.05--0.1 \\
$^3F_3$& 2.23  &0.15 &2.22&0.17&  2.20--2.26 & 0.1--0.2 \\
\hline
\end{tabular}\end{center}
\end{table}

%By using the developed model, we can also estimate the elastic
%($NN$) part of the total dibaryon width by simply comparing the
%values of $\Gamma_0$ and $\Gamma_{\rm th}$. Thus, the decay width
%into the $NN$ channel turns out to be less than 20\% of the total
%width for all three partial waves considered. This result agrees
%qualitatively with the previous
%estimates~\cite{Arndt,Strak,Strak91} which typically give the $D
%\to NN$ partial width to be 10--20\% of the total width for the
%isovector dibaryons.

Thus, in this section we have got a very good quantitative
approximation for the real parts of the partial phase shifts in
the $NN$ channels considered and a rather good qualitative
approximation for the imaginary parts of these phase shifts in a
broad energy range starting from zero energy which is very far
from the position of the ``bare'' dibaryon. The crucial point of
the results presented above is an agreement of the masses and
widths of the dressed dibaryons obtained by fitting the $NN$ phase
shifts in our model with the parameters of experimentally found
dibaryons.

\section{Channel $\bm{^1S_0}$. Description of the repulsive core effects}
\label{res2} One of the main ingredients of the conventional
models for $NN$ interaction is a well-known short-range repulsion
induced by vector-meson exchange. However, from the modern point
of view such a mechanism looks doubtful.\footnote{See, e.g.,
Ref.~\cite{Barnes} where the authors claim: ``A literal
attribution of the short-range repulsive core to vector meson
exchange, as opposed to a phenomenological parametrization, of
course involves a {\em non sequitur}: since the nucleons have
radii $\approx$ 0.8 fm and the range of the vector exchange force
is $\hbar/m_\omega c\approx 0.2$~fm one would have to superimpose
the nucleon wavefunctions to reach the appropriate internucleon
separations. The picture of distinct nucleons exchanging a
physical $\omega$-meson at such a small separation is clearly a
fiction\ldots''} We will demonstrate below how the repulsive core
effects can be reproduced within the framework of the quark and
dibaryon models (see also Refs.~\cite{Faess,Sazonov,YAF2013}).

\begin{figure}[h!]
\centering\epsfig{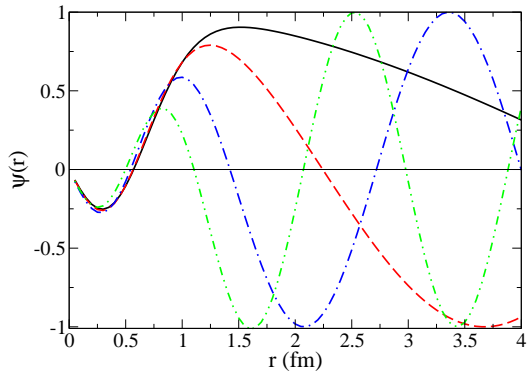}
\caption{\label{fig5} Nodal behavior of scattering functions in
the $^1S_0$ partial wave at different energies $T_{\rm lab}$: 10
MeV (solid curve), 100 MeV (dashed curve), 500 MeV (dash-dotted
curve), and 1 GeV (dash-dot-dotted curve).}
\end{figure}

As is well known, the repulsive core effects are manifested very
clearly in the channel $^1S_0$, where the phase shifts become
negative already at collision energies $T_{\rm lab} \simeq
250$~MeV. The dibaryon model \cite{JPhys2001,AnnPhys2010} predicts
for the $S$-wave $NN$ interaction the dominating six-quark
configuration of the type $|s^4p^2[42]_xLST\rangle$ with a
two-quantum ($2\hbar\omega$) excitation. When projecting this
$2\hbar\omega$-excited configuration onto the $NN$ channel, the
$NN$ relative-motion wavefunction $\psi(r)$ automatically acquires
an internal node. This node is rather stable and its position
$r_{c}$ moves only weakly with increase of collision energy (see
Fig.~\ref{fig5}). Moreover, the node position ($r_{c} \sim
0.5$~fm) turns out to be very close to that of the traditional
repulsive core.

In the dibaryon model \cite{JPhys2001,AnnPhys2010} appearance of a
{\em stationary node} in the $S$-wave $NN$ interaction is provided
by an orthogonality condition between the fully symmetric and the
mixed-symmetry six-quark configurations, which is realized by
adding a projecting operator to the $NN$ potential:
\begin{equation}
V_{\rm rep}=\lam|\phi_0\rangle\langle \phi_0|, \label{v_rep}
\end{equation}
where $\lam \to \infty$, and $|\phi_0\rangle$ is a projection of
the fully symmetric wavefunction $|s^6[6]\rangle$ onto the $NN$
channel. The idea behind inclusion of this term into the $NN$
potential is that the six-quark configuration $|s^6[6]\rangle$
gives a much smaller contribution to the $S$-wave $NN$ interaction
than the mixed-symmetry component. Hence, in case of $S$- and
$P$-wave channels which exhibit strong repulsive core effects
($^3P_0$, $^3P_1$ and $^1P_1$), the effective $NN$ Hamiltonian
(\ref{Heff}) should be supplemented with the orthogonalizing
pseudopotential (\ref{v_rep}) with a large positive coupling
constant $\lam$.\footnote{For the $P$ waves, $|\phi_0\rangle$ is a
projection of the six-quark wavefunction $|s^5p[51]_xLST\rangle$
onto the $NN$ channel.}

Fortunately, this modification does not add to our model any free
parameters, so that, the number of adjustable parameters for the
bare dibaryon remains the same. Thus, varying the position of only
one complex pole in the $^1S_0$ channel, one can achieve a very
reasonable description of the phase shifts in a wide energy range.
However, the resonance in this channel occurs to be very broad,
hence, few different sets of parameters can be used to fit the
phase shifts. In Fig.~\ref{fig5} we show the results for the
potential parameters $\lam_1=1.184$ GeV, $r_0=0.51$ fm, the
initial dibaryon mass $M_0 = 2.364$~GeV, and the initial width
$\Gamma_0 = 0.044$~GeV with $l_\pi=1$, $l_{NN}=0$ and $\Lam = 0.4$
GeV/$c$.

\begin{figure}[h!]
\centering\epsfig{file=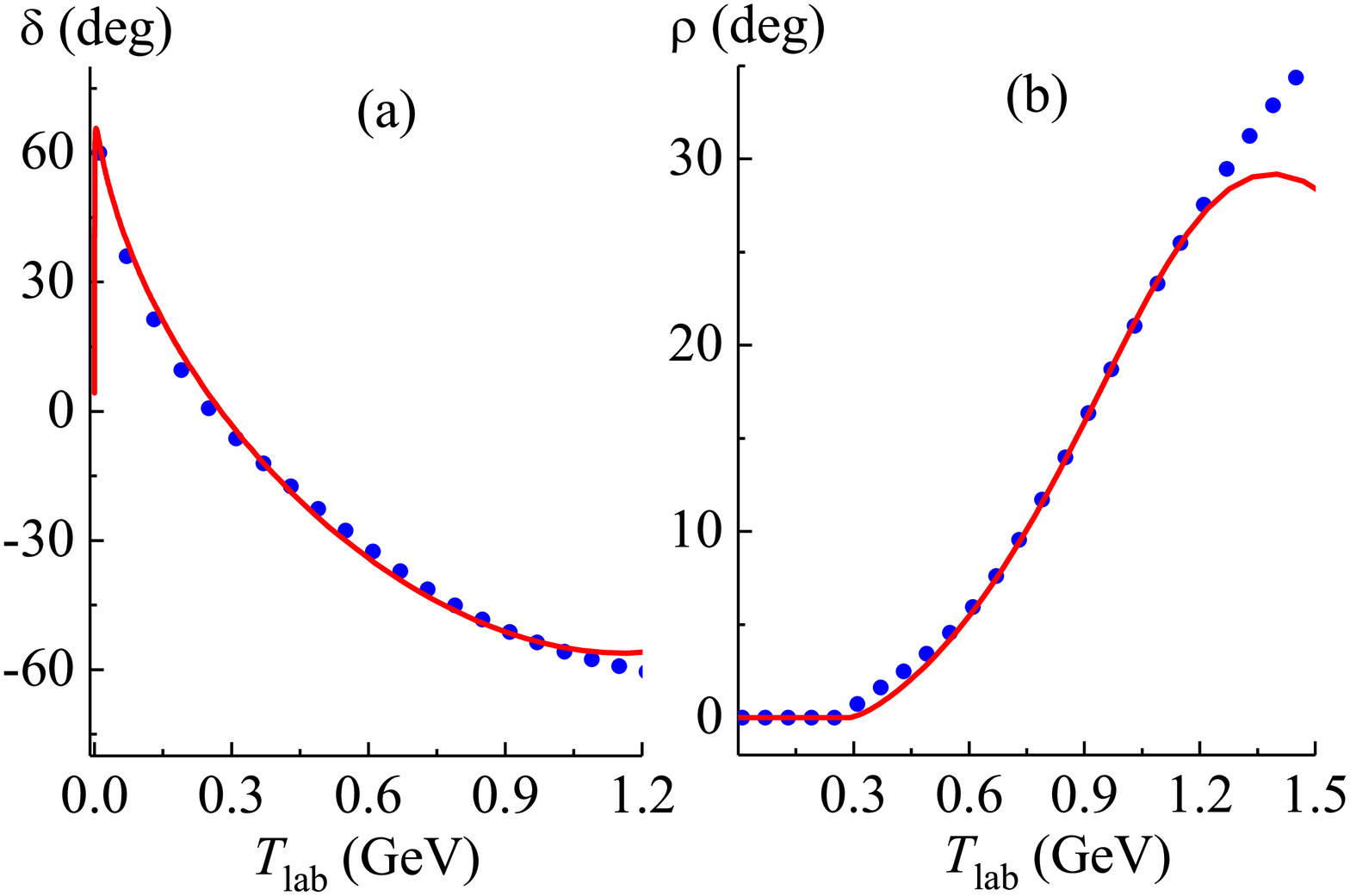,width=\columnwidth}
 \caption{\label{fig6} The same as in Fig.~\ref{fig2} for the $^1S_0$ partial wave.}
\end{figure}

It should be stressed that the total $S$-matrix in the $^1S_0$
channel has two poles: the first corresponding to the well-known
singlet deuteron just above the $NN$ threshold (at $E=-0.067$ MeV
on the unphysical complex energy sheet) and the second being a
broad high-lying resonance with $M_{\rm th}=2.59$ GeV and
$\Gamma_{\rm th}=0.63$ GeV. We emphasize here once again that the
mass and width of the bare $^1S_0$ dibaryon were taken to be the
same for description of both real and imaginary phase shifts.

These results give a strong indication for existence of a second
high-lying dibaryon in the $^1S_0$ channel in addition to the
near-threshold state (the singlet deuteron) predicted by Dyson and
Xuong \cite{Dyson} many years ago. It is interesting to note that
while the second pole for the bare dibaryon has been postulated
when constructing the effective potential (\ref{Heff}), the first
one appears in the course of dressing, i.e., when the coupling
between the $NN$ and dibaryon channels gets switched on.

\section{Conclusion}
\label{concl}

We have demonstrated in this work that the hypothesis about the
dibaryon origin of the basic nuclear force makes it possible to
describe properly the behavior of both real and imaginary parts of
$NN$ phase shifts (at least in some partial-wave channels
characterized by a large inelasticity) in a rather wide energy
range 0--600 MeV with only a few basic parameters for the initial
``bare'' dibaryon. This result should be compared to the fact that
the traditional realistic $NN$ potentials can describe only the
real $NN$ phase shifts in the energy range 0--350 MeV.

What is even more interesting, the results of the present study
can also explain the strong partial-wave dependence of the $NN$
phase shifts, which must depend upon the dynamics of dibaryon
states in the given partial channel, their spin, isospin, parity,
etc. In turn, the dibaryon dynamics is determined completely by
the QCD degrees of freedom. Without taking into account the
intermediate dibaryons, the $NN$ interaction potential must have
very complicated and non-transparent structure and adopt numerous
terms with ${\bf S \cdot L}$, $L^2$, etc., operators to describe
the empirical channel dependence of the $NN$ phase shifts.

{\bf Acknowledgment.} The work has been partially supported by RFBR,
grants Nos. 19-02-00011 and 19-02-00014.

\end{document}